\documentclass[%
superscriptaddress,
preprintnumbers,
 amssymb,
 aps,twocolumn,
prb,showkeys,
floatfix,
]{revtex4-1}

\usepackage{graphicx}
\usepackage{dcolumn}
\usepackage{bm}
\usepackage{amsmath,environ,mathtools}
\usepackage{makecell}
\usepackage{amssymb, textcomp}
\usepackage{caption}
\usepackage[T1]{fontenc}

\usepackage{booktabs,caption}
\usepackage[flushleft]{threeparttable}
\usepackage{gensymb}
\usepackage{booktabs,siunitx}
\usepackage{amsmath}

\usepackage{xcolor}

\begin{document}


\title{\textit{Ab initio} investigations of point and complex defect structures in B2-FeAl}

\author{Halil \.{I}brahim S\"ozen}
\affiliation{ Max-Planck-Institut f\"ur Eisenforschung GmbH, Max-Planck-Str.~1, 40237 D\"usseldorf, Germany}
\author{Tilmann Hickel}
\affiliation{ Max-Planck-Institut f\"ur Eisenforschung GmbH, Max-Planck-Str.~1, 40237 D\"usseldorf, Germany}
\author{J\"org Neugebauer}
\affiliation{ Max-Planck-Institut f\"ur Eisenforschung GmbH, Max-Planck-Str.~1, 40237 D\"usseldorf, Germany}




\date{\today}

\begin{abstract}

In this work we have studied  the defect structure and corresponding defect concentration investigations through the theoretical, experimental and computational works on B2-type Fe-Al alloys. We have used \textit{ab initio} framework in order to investigate the defect structure. To have a proper explanation for high defect concentration in B2-FeAl, we did not confine with point defect, but extend the work on defect complexes. The possible defect formation energies were calculated with the dependence of chemical potential and carefully investigated against supercell size and the effect of magnetism. The calculations revealed that the double Fe antisite at Fe rich condition, the single Fe vacancy at intermediate region (i.e in the stoichiometry) and the double Al antisite is the dominant defect close to Al rich condition, where mainly Al rich region was unstable. From the obtained defect formation energies, defect concentrations were calculated at different temperatures with respect to Al concentration for B2-FeAl. It has been found that increasing Al content and temperature gradually leads to increase in the vacancy content. It has also seen that the dominant defect for all temperature ranges was the single Fe vacancy at the exact stoichiometry and the highest single Fe vacancy content detected with 1.6 \% at 1450 K.

\end{abstract}

\pacs{Valid PACS appear here}
\maketitle


\section{Introduction}

Fe-Al alloys have been a field of interest in material science since the 1930s. It has been a promising candidate for industrial applications once it was discovered that Fe-Al alloys have high corrosion and sulphidation resistance properties \cite{Tortorelli1992} with composition of Al more than 20 at.~\%, compared to steels and Fe-based commercial alloys. It has also been realized that alloys with this composition have lower density than stainless steels \cite{Jordan2003} and have comparable tensile strength as ferritic and austenitic steels \cite{Xiao1995}. These properties make Fe-Al alloys attractive for the industry, where inexpensive and high temperature structural materials are used. 

However, the extensive technical applications of iron aluminades are constrained due to the low ductility at ambient temperatures and poor fracture toughness \cite{Deevi1996,Liu1995}. The thermomechanical properties of high temperature intermetallics are closely related to the defect structure and their migration, particularly comparing to other intermetallics B2-FeAl shows a very high vacancy concentration of several percent at elevated temperatures \cite{Ho1978,Y.A.Chang1993,Haraguchi2003,Kogachi1997,Kogachi1998}. Therefore, in order to develop more ductile Fe-Al alloys, which depend on the deep understanding of their deformation mechanisms that linked to the defect structure, and to estimate the high temperature mechanical behavior a physical understanding of atomistic defect formation, concentration and migration is fundamentally required.

In this work, we focused on defect formation and concentration only. Fe-Al alloys form many different phases, especially in the Al-rich region, but of technological interest are mainly the DO$_3$-Fe$_3$Al and the B2-FeAl phase. In particular to high temperature applications, B2-FeAl is preferred due to the high melting point and wide range of stability. Therefore, in this paper we will mainly consider the letter.

Defect and diffusion behavior of Fe-Al alloys have been studied experimentally \cite{Jordan2003,R.Kerl1999,J.Wolff1999,Hanc2009,Nakamura2003,Eggersmann2000} and theoretically  \cite{Mayer1995,J.Mayer1999,C.L.Fu1993,R.Drautz1999,Kellou2006,Amara2010}. Fu \textit{et al.} \cite{C.L.Fu1993} performed \textit{ab initio} calculations for the binding energies to make di-vacancy in stoichiometric B2-FeAl. They reported a significantly high binding energy with a value of 0.57 eV, which indicates that there is a strong tendency for vacancy clustering and vacancies can be annealed out to be open structures such as dislocations, voids or grain boundaries. Haraguchi \textit{et al.} \cite{Haraguchi2005} confirmed this with as spun and annealed ribbons and fully annealed powder sample experiments. The group of F\"{a}hnle used grand canonical approach and published several extended studies \cite{Mayer1995,J.Mayer1999,R.Drautz1999,J.Mayer1997}. One of their main results is that there is no Al vacancies in B2-FeAl. Other authors also confirm that result.

However, there are wide spread of the reported defect formation energies in B2-FeAl. As an example, there are many results of Al vacancy formation energies, which are changing from 1.62 eV \cite{Besson2006} (with Boron impurity) to 4.69 eV \cite{Kellou2004}. In addition to this, these authors conclude that vacancy clustering is negligible that is in contrast to results of Fu and Haraguchi. In this work, it will be shown that the main aspects for that much of wide range (as in example of Al vacancy) defect formation energies can rise due to the three reasons: ($i$) unconverged supercells, ($ii$) non-magnetic Fe and ($iii$) calculations that are independent from chemical potential.

Therefore, in this paper we had performed a systematic density functional theory (DFT) calculations based on criteria above to have a proper explanation for the defect structure and concentration. The paper is organized as follows: In Sec.~\ref{Sec_CompDets} we give a brief explanation on computational details and methodology. In Sec.~\ref{Sec_FormEne}, we describe how to calculate the considered defect formation energies. The comparison of the calculated defect concentrations against experiments is discussed in Sec.~\ref{Sec_Def_Conc}. We finally conclude the paper with remarks and discussions in Sec.~\ref{Sec_Conc}.

\begin{table*}[ht]
\center
\label{table:1}
\caption{Comparison of the calculated results against experimental data for the bulk B2-FeAl. The lattice parameter $a$ and bulk modulus $B$ dertermined by a fit to the Murnaghan equation of state\cite{Murnaghan1944}. \textit{H} denotes to the formation enthalpy/atom.} 
\begin{tabular}{  l  c  c  c  c  c }
    \toprule
     & \textit{a} (\AA) & \textit{B} (GPa) & Cohesive energy (eV) & \textit{H} (eV) & Magnetic moment $(\mu_{B})$ \\ 
    \midrule    
    Nonmagnetic & 2.870 & 179.7 & -6.343 & -0.315  & 0 \\ 
    \midrule
    Ferromagnetic & 2.874 & 175.3 & -6.359 & -0.331 & 0.71 \\
    \midrule
    Experiment.\cite{Boer1988,Villars1988} & 2.86, 2.88 & 152 & N.A & -0.33 & 0 \\
    \bottomrule
\end{tabular}
\label{Tab_BulkProps}
\end{table*}

\section{Computational Details}
\label{Sec_CompDets}

In this work, all first-principles calculations are performed in the framework of DFT with the Vienna  \textit{ab initio} Simulation Package (VASP) \cite{G.Kresse1996,Kresse1996}. In this package, we use the projector-augmented wave method (PAW) to describe the ion-electron interactions. Exchange-correlation is treated within the generalized gradient approximation (GGA) of Perdew-Burke-Ernzerhof (PBE)\cite{Perdew1996}, which in general gives more reliable results for transition metal systems compared to the local density approximation (LDA).

As mentioned in the introduction, one of the reasons for the large deviations in the result of previously performed simulations was the usage of unconverged supercells, meaning too small supercells. In order to  ensure the satisfactory convergence with respect to supercell size, bulk and defect simulations are performed with 16 atom (2$\times$2$\times$2), 54 atom (3$\times$3$\times$3) and 128 atom (4$\times$4$\times$4) supercells. It has been determined to use 128 atom supercell for defect structure investigations, where defect formation energies are converged. 

The second reason for the deviating results was non-magnetic (NM) calculations. Therefore, calculations are performed in spin-polarized scheme, meaning iron is assumed to be ferromagnetic (FM). The ferromagnetic and non-magnetic iron affects the defect formation energies especially for the Al vacancy defect formation energy (the detailed explanation will be given in Sec.~\ref{Sec_FormEne}). According to theory, bulk B2-FeAl is ferromagnetic. However, experiments show that B2-FeAl is paramagnetic at least down to 1 K \cite{G.R.Caskey1972}. That may be due to the current DFT functionals \cite{Mohn2001} and there are also reports of Mohn \textit{et al.} \cite{Mohn2001} that he overcome that disagreement by taking the correlation effects into account within the LDA+U framework. Furthermore, Smirnov \textit{et al.} \cite{Smirnov2005} also proposed that the energy differences between the ferromagnetic and paramagnetic state are almost negligible in the concentration range that we are interested. Also in between point defects Fe antisite is the dominant one in the Fe rich region, that will be explained in later sections, leads to net magnetic moment \cite{Zaroual2000a}. For these reasons, all simulations in this work will be performed in the ferromagnetic state and the energy of the ferromagnetic state is considered to be the ground state of the system. The third reason for the deviating results, which is excluding chemical potential dependence will be mentioned in detail in Sec.~\ref{Sec_FormEne}.

As the $k$-point sampling the Monkhorst-Pack \cite{Monkhorst1976} method has been used. For the 16 atoms supercell a 12$\times$12$\times$12 mesh, 54 atoms case 8$\times$8$\times$8 and for 128 atoms case 6$\times$6$\times$6 $k$-points mesh density was applied. After some set of convergence tests, it is determined to use 420 eV as cut-off radius. The smearing parameter was selected as 0.15 eV. The number of valance electrons taken into account was 8 and 3, for Fe and Al, respectively.  A dense mesh of $k$-points and considered parameters satisfied an energy convergence of 1 meV/atom, which was considered as the convergence criterion.

\section{Defect Formation Energies}
\label{Sec_FormEne}

In order to investigate defect properties, reliable reference energies from bulk defect-free calculations are crucially important. Therefore, bulk calculations have been performed for defect-free B2-FeAl structure with the ferromagnetic and non-magnetic state. Lattice constant, bulk modulus, cohesive energy (energy per atom), formation enthalpy and magnetic moment are found for different magnetic phases and compared with experiments given in Tab.~\ref{Tab_BulkProps}. 

The agreement of the theoretical lattice constant and the formation enthalpy with the experimental results is rather promising. However, the bulk modulus seems to be slightly over estimated. This looks like a shortcoming of DFT, but in upcoming sections it will be clarified that this difference can be explained in a straightforward way by taking into account defects. The theory tells that the Fe local moments order ferromagnetically in B2-FeAl is also confirmed by slightly lower energy/atom value in ferromagnetic state. Nevertheless, it should be noted that the lattice constant values and other quantities are remarkably similar to each other, which can be considered that the structure is actually not very ferromagnetic.

\begin{figure}
\begin{center}
\includegraphics[width=0.45\textwidth]{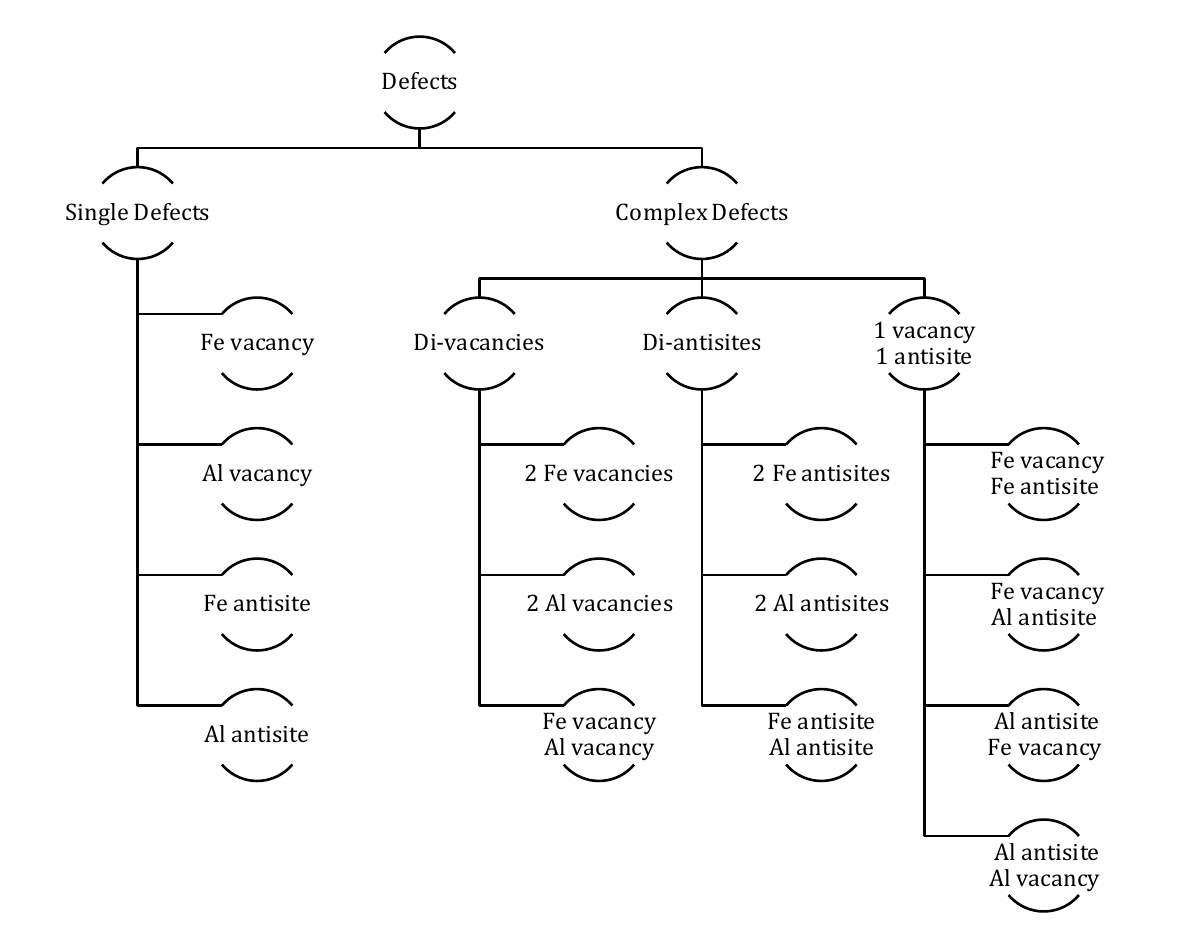}
\caption{Considered single and complex defects in B2-FeAl for our theoretical investigations.}
\label{Fig_ListofDefects}
\end{center}
\end{figure} 

In B2-FeAl compound there are four kinds of native defects, which are Fe vacancy, Al vacancy, Fe antisite and Al antisite (due to the size of the atoms no interstitials are observed). To be able to explain the high amount of defect (especially vacancy) concentration in B2-FeAl, we also add the possible complex defects. All considered defects are listed in Fig.~\ref{Fig_ListofDefects}. The defect complexes with two atoms are: ($i$) di-vacancies, ($ii$) di-antisites, ($iii$) one antisite one vacancy. 

As listed in the introduction section, that the last reason for the diversity in the defect formation energies is using different reference states instead of taking into account chemical potential. In this work, the chemical potential is used as an adjustable parameter throughout all expressions for the defect formation energy and solution enthalpy. 
The chemical potential of Fe and Al are not independent of each other since both species are in equilibrium with B2-FeAl. This  is expressed by the following formation enthalpy formula:

\begin{equation}
\mu_{\mathrm{Fe}} + \mu_{\mathrm{Al}} = H_{\mathrm{FeAl}}^f,
\label{eq:Equ1}
\end{equation}

\noindent where $H_{\mathrm{FeAl}}^f$=-0.662 eV is the formation enthalpy of bulk B2-FeAl (compare with the Tab.~\ref{Tab_BulkProps}). Thus only one of the chemical potentials can be changed independently, which will obviously change the defect formation energies and the composition of the alloy as well. In this work $\mu_{\mathrm{Fe}}$ has been chosen as the independent chemical potential and we plot the defect formation energies with respect to $\mu_{\mathrm{Fe}}$. It still cannot be arbitrarily changed, since the condition $H_{\mathrm{FeAl}}^f<\mu_{\mathrm{Fe}} <0$ has to be fulfilled to ensure the stability of the system. By definition $\mu_{\mathrm{Fe}}=0$ corresponds to the chemical potential of bulk Fe. The B2-FeAl can only be stable if $\mu_{\mathrm{Fe}}$ lies between those two values, otherwise it becomes energetically more favorable that the system segregates into pure Fe or Al.

\subsection{Single Defect Formation Energies}

The formation of a defect costs energy, which can be calculated within DFT. The formation energy of an $A$ vacancy in an $AB$ compound is given by: 

\begin{equation}\label{eq:Equ2}
E_{vac,A}^{f}=
E(A_{N-1}B_{N}) + \mu_{A} - N E(AB),
\end{equation}

\noindent where $N$ is the number of atoms and $A (B)$ stands for either Fe or Al. $E(AB)$ is the energy of an $AB$ unitcell,  $E(A_{N-1}B_{N})$ is the energy of a $2N$ supercell with an $A$ atom removed and $\mu_{A}$ is the chemical potential of the $A$ species. Similarly the $A$ antisite formation energy can be defined as: 

\begin{multline}\label{eq:Equ3}
E_{AS,A}^{f}=\\
E(A_{N+1}B_{N-1}) + \mu_{B} - N E(AB) - \mu_{A},
\end{multline}

\noindent $E(A_{N+1}B_{N-1})$ represents the energy of a $2N$ supercell, where one $B$ atom is replaced by an $A$ atom. This is what we call an $A$ antisite atom. 

As it seen from the definition of the formation energies, they will also depend on the size of the supercell. Thus convergence with respect to supercell size has to be carefully checked. Therefore, some set of our simulations are  performed  for 16, 54 and 128 atoms supercells. In each of the supercells only one defect is considered. This corresponds to a defect concentration of 6.25~\% for the 16 atoms supercell, 1.85~\% for the 54 atoms supercell and 0.78~\% for the 128 atoms supercell. For each defect that considered in different sizes of supercell we calculate the lattice constant and bulk modulus. 

\begin{figure}[h!]
\begin{center}
\includegraphics[width=0.45\textwidth]{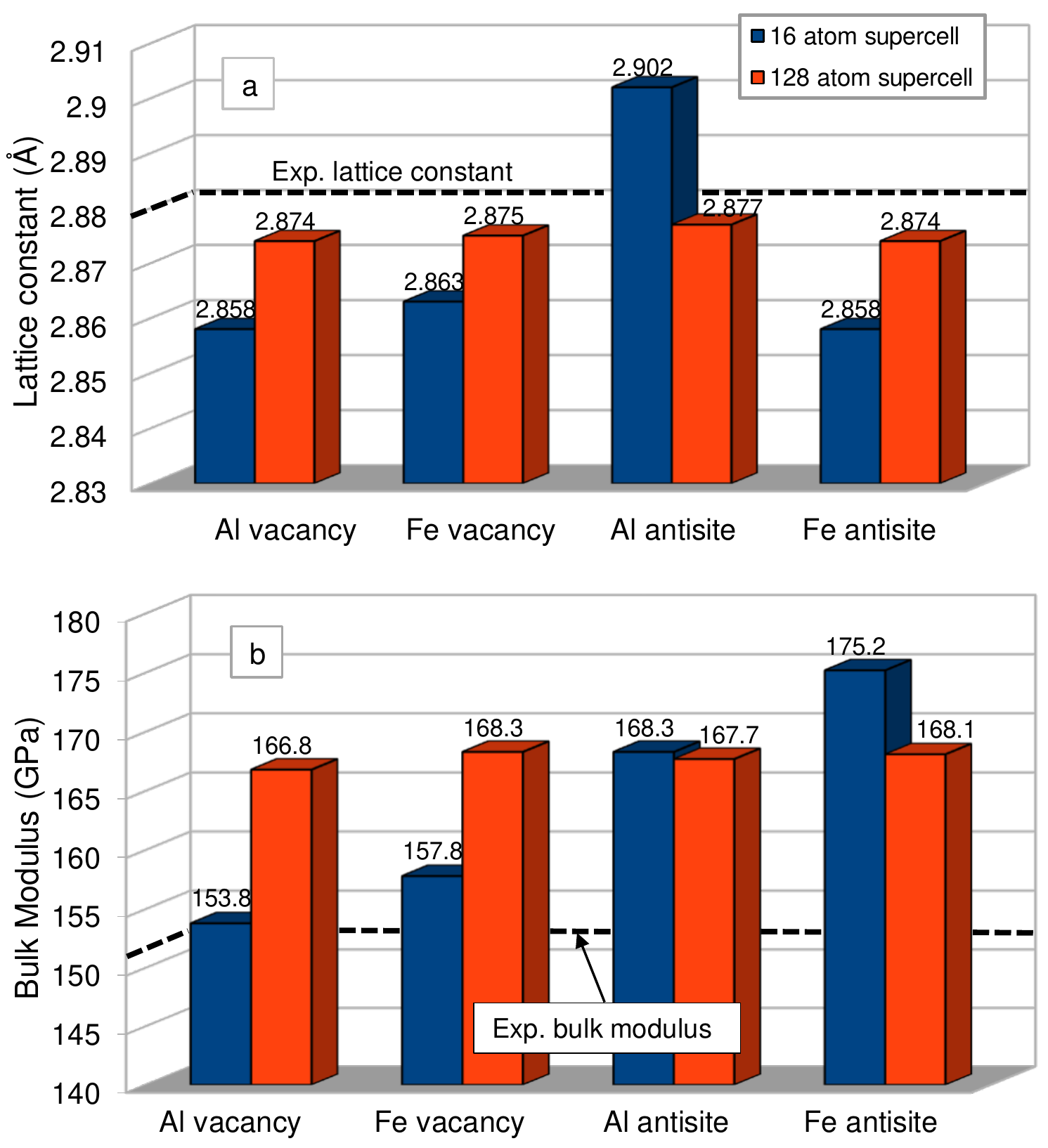}
\caption{\textit{Ab initio} calculations of lattice parameters (a) and bulk modulus (b) of B2-FeAl including single defects for 16 and 128 atom supercells. The dashed lines indicate the corresponding experimental values \cite{Villars1988}.}
\label{Fig_lattConstANDBulkMods}
\end{center}
\end{figure}

As it can be seen from Fig.~\ref{Fig_lattConstANDBulkMods}, the lattice constant (a) and bulk modulus (b) are quite supercell size dependent and it changes from 16 atom supercell to 128 atom supercell. The lattice constants of 16 atom supercell calculations are underestimated as compared to the 128 atom calculations, except for the Al antisite. That is quite understandable, because the atomic radius of Al is larger than that of the Fe and in a 16 atom supercell with an Al antisite, which means 9 Al atoms and 7 Fe atoms that effect can be seen significantly. However, for the 128 atoms supercell calculations that effect becomes negligible and the values are close to the bulk value for all the considered defects.

On the other hand, the bulk modulus agrees much better with the experimental values of 152 GPa once the single defects are added. Note that, it was  calculated for a pure system was 179.7 GPa for the non-magnetic case and 175.3 GPa for the ferromagnetic case (see from Tab.~\ref{Tab_BulkProps}). Therefore, one can explain the apparent failure of the theory to predict the experimental values that is mentioned in the sections above. It is not possible to produce a defect-free B2-FeAl, which means that there will be always vacancies. Especially, the experiments that are done in the non-stoichimetric regime (Al content between 40 at.~\% and 50 at.~\%) have to contain a rather large concentration of structural defects. If one takes those vacancies into account the agreement between theory and experiment becomes much better. 

\begin{figure}[h!]
\begin{center}
\includegraphics[width=0.5\textwidth]{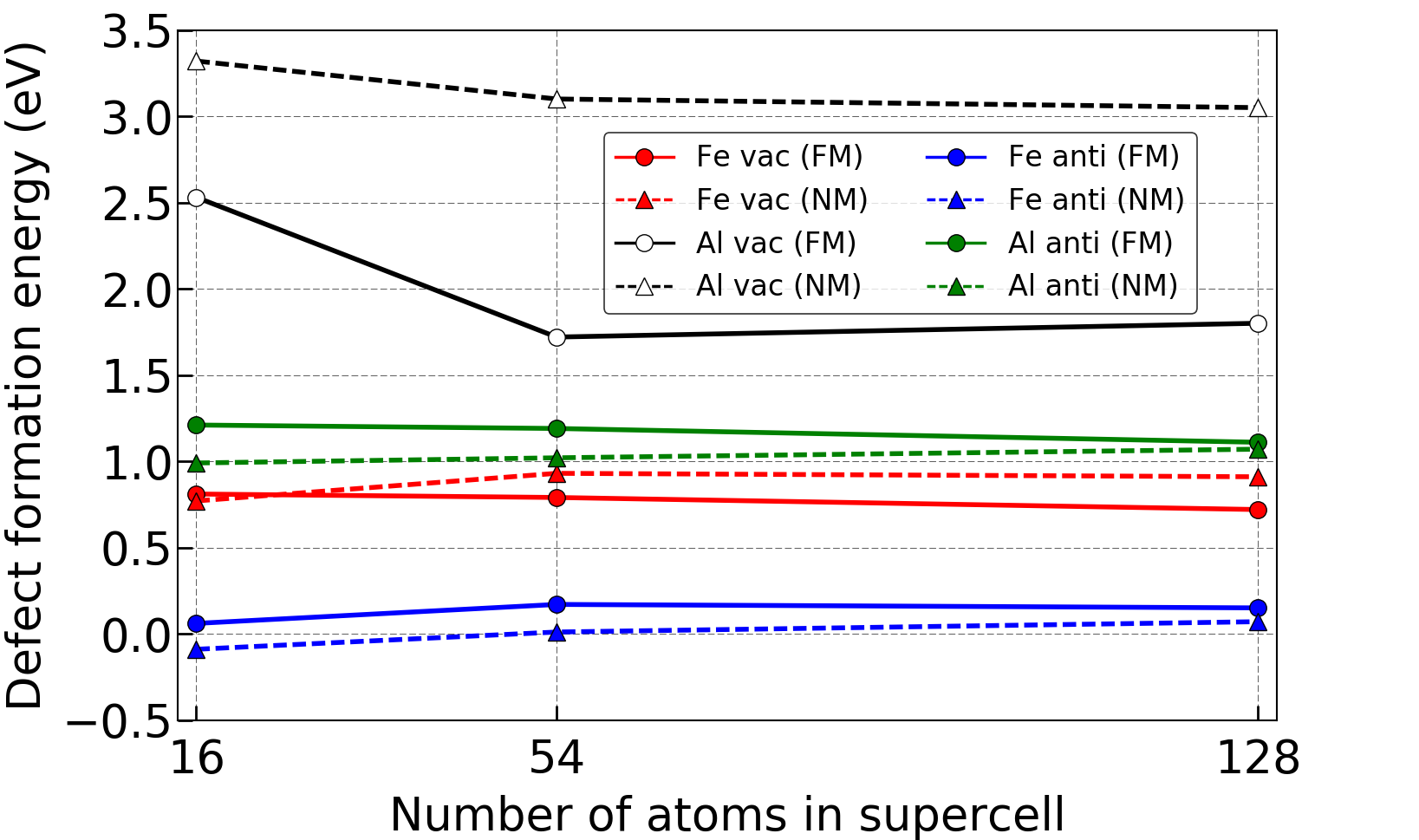}
\caption{\textit{Ab initio} calculations of the single defect formation energies in ferromagnetic and non-magnetic state with respect to changing supercell size. All of the defect formation energies are plotted with respect to the Fe rich conditions.}
\label{Fig_MangnetismAndSizecell}
\end{center}
\end{figure}

In Fig.~\ref{Fig_MangnetismAndSizecell}, the supercell size and the magnetism effect on the defect formation energies are investigated. The Al vacancy formation energy changes quite drastically from 2.53 eV to 1.80 eV, i.e. by about 40~\%, when changing the supercell size from 16 to 128 atoms for spin polarized stoichiometric conditions. This is surprising since the difference for the other defects are much smaller, which indicates that 16 atom supercell is not yet converged. Nevertheless, in the case of 54 atom cell size the deviation to the 128 atom supercell is only 4.4~\%. Here, the Al vacancy formation energy for the 54 atoms case is 1.72 eV. Thus the significant change of the ferromagnetic formation energy has to be due to the magnetism. We also performed non-spin polarized calculations and found that the defect formation energy in this case changes by only 8.8~\% from 3.32 eV (16 atoms case) to 3.05 eV (128 atoms case). At the same time this explains why there is a large deviation of Al vacancy formation energies reported in the literature. Especially in older publications and reports smaller supercells were used due to computational cost and often magnetism was neglected as well. Thus the reported formation energies were much too high. The Fe vacancy does not show magnetism dependence as much as Al vacancy. Nevertheless, FM and NM 128 atoms supercell formation energies 0.72 eV and 0.91 eV, respectively.
 
In the case of single antisite defects, the Al antisite calculations for both spin polarized and non-spin polarized are well converged compared to the 128 atom supercell, i.e. does not depend on supercell size. In addition, Fe antisite calculations for spin polarized and non-spin polarized calculations converged well, where for spin polarized calculations antisite defect formation energy for 54 atoms case 0.17 eV and for 128 atoms case 0.15 eV. 

\begin{figure}[h!]
\begin{center}
\includegraphics[width=0.45\textwidth]{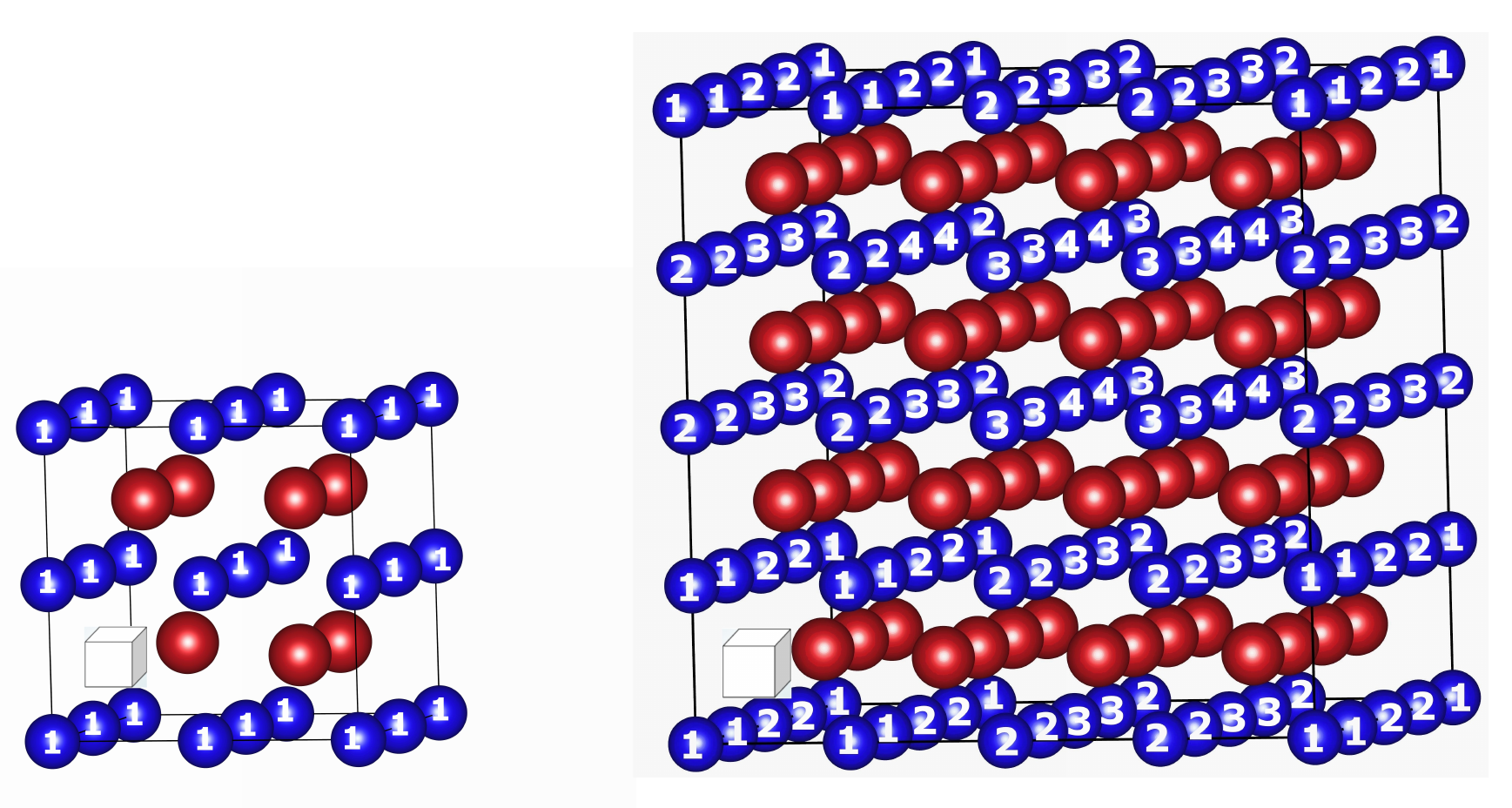}
\caption{Schematic representation of 16 atom (left) and 128 atom (right) supercells with an Al vacancy. The Al vacancy is symbolized by the white box. Numbers on Fe atom indicates the nearest neighbor shell to the Al vacancy. Note that periodic boundary condition is considered in the illustration.}
\label{Fig_Bader}
\end{center}
\end{figure}

\begin{figure}[h!]
\begin{center}
\includegraphics[width=0.5\textwidth]{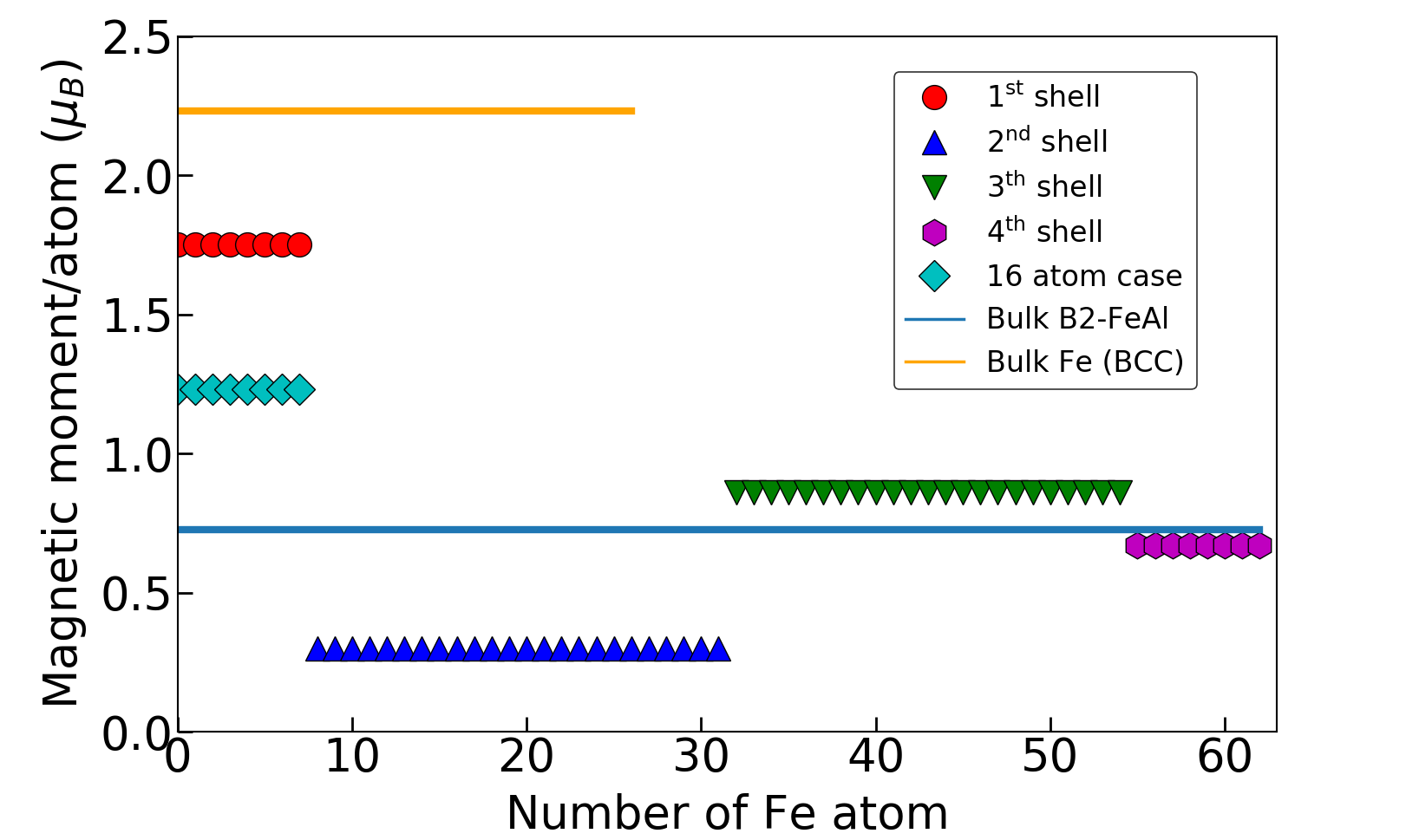}
\caption{Calculated magnetic moments of Fe atoms in 16 and 128 atom supercells with one Al vacancy included. Both crystal structures are shown schematically in Fig.~\ref{Fig_Bader}.}
\label{Fig_MagMomChart}
\end{center}
\end{figure}

As discussed before that B2-FeAl is slightly ferromagnetic at 0 K and paramagnetic at higher temperatures. Nevertheless, experiments showed that \cite{G.R.Caskey1972,Zaroual2000a,Domke1984}, an Fe atom in Al sublattice or an Al vacancy, which surrounded by Fe atoms can exhibit an effective magnetic moment. Bester \textit{et al.} \cite{G.Bester2002} also states that, inclusion of the spin polarization supports the formation of Fe antisite atoms because of the gain of interatomic exchange energy. Therefore, in order to investigate the effect of magnetism in a more detailed way, the individual magnetic moment of each atoms is determined by the Bader analysis \cite{Henkelman2006}. Al alone is a non-magnetic element, but induced some magnetic moments (-0.17 $\mu_B$ to -0.24 $\mu_B$ ) due to the presence of Fe atoms. However, the magnetic moments of the Fe atoms are rather complex and depend on the distance between the respective atom and the defect. Spin polarized calculations with 16 and 128 atom supercells and an Al vacancy were performed. Figure \ref{Fig_Bader} shows the corresponding supercells and the Al vacancy and Fig.~\ref{Fig_MagMomChart} reveals the calculated magnetic moments. 

For the 16 atom supercell all the Fe atoms are nearest neighbors (NN) of the Al vacancy, which makes their magnetic moments of all the same and equal to 1.23 $\mu_B$. There is no change in the magnetic moments neither decreasing nor increasing, therefore one can also expect that energy will not change as a function of the magnetic moments.
The dependence of magnetic moments to distance can be seen clearly in the case of 128 atom suprecell, where 1$^{\mathrm{st}}$, 2$^{\mathrm{nd}}$, 3$^{\mathrm{rd}}$ and 4$^{\mathrm{th}}$ shell has 1.75 $\mu_B$, 0.32 $\mu_B$, 0.86 $\mu_B$ and 0.69 $\mu_B$ magnetic moment. As shown in the Fig.~\ref{Fig_MagMomChart} there is a Friedel-like oscillatory behavior for magnetic moments. In a way that once an Al atom is removed, all NN atoms are Fe, that are labeled with integer 1 in Fig.~\ref{Fig_Bader}, and when the structure is relaxed these atoms come closer and form a cluster with 8 Fe atoms that gives a high magnetic moment of 1.75 $\mu_B$. The NN Fe atoms magnetic moments are coming close to bulk body centered cubic (BCC) Fe atoms magnetic moment that is indicated by the orange line, which is 2.20 $\mu_B$ and 4$^{\mathrm{th}}$ NN magnetic moments of Fe atoms is 0.69 $\mu_B$, which is close to the bulk B2-FeAl magnetic moment value 0.71 $\mu_B$ that is indicated by the blue line. It is also a clear sign that the selection of a 128 atoms supercell is magnetically converged and these 4$^{\mathrm{th}}$ shell atoms will not be significantly affected by the vacancy.

\begin{figure}[h!]
\begin{center}
\includegraphics[width=0.48\textwidth]{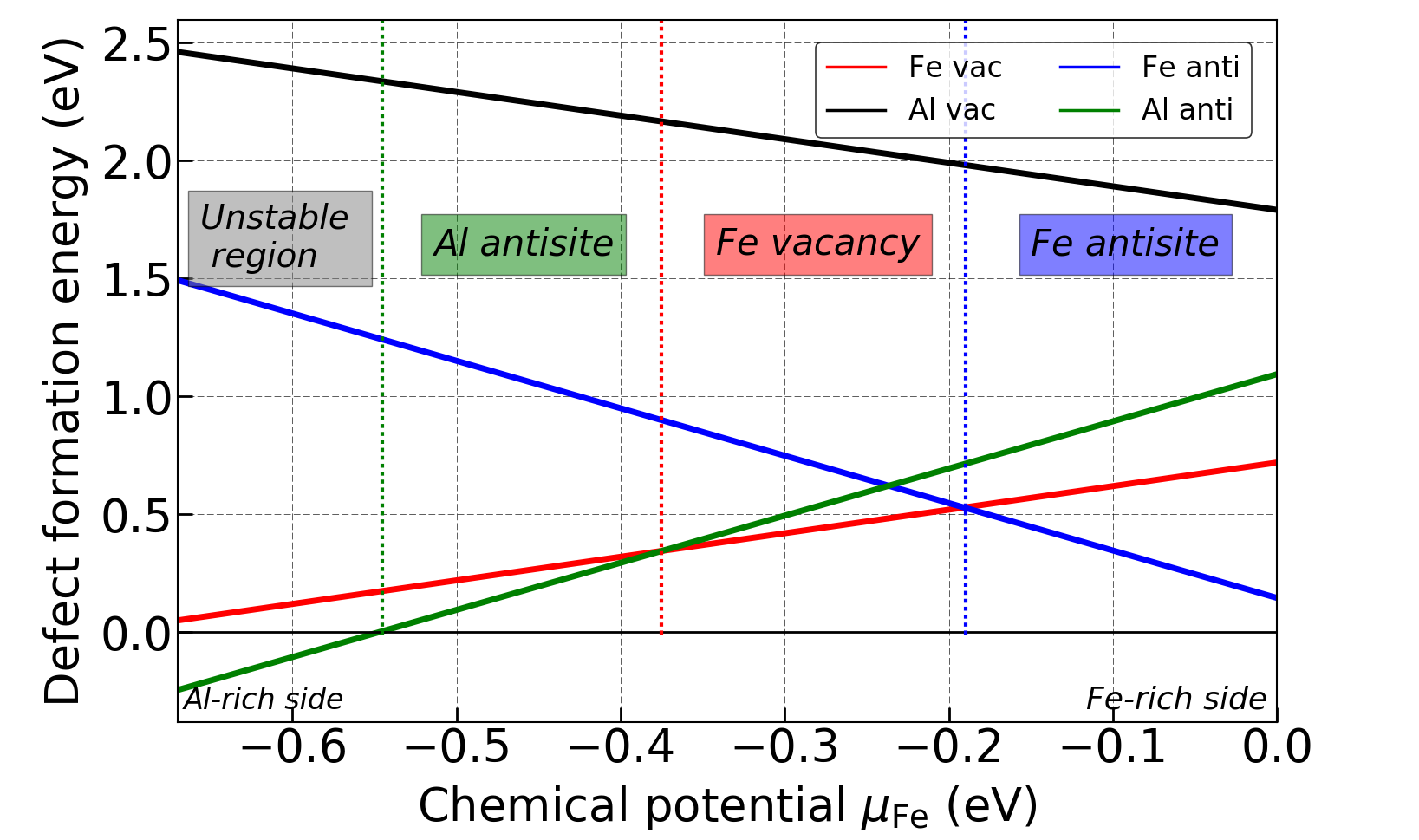}
\caption{Calculated single defect formation energies with the dependence on $\mu_{\rm{Fe}}$ the for the spin polarized 128 atom supercells. The left hand side of the plot corresponds to Al rich conditions and right hand side Fe rich. Vertical lines represent the dominant defect regions.}
\label{Fig_SingleDefects}
\end{center}
\end{figure}

The defect formation energies are calculated according to Eqs.~\ref{eq:Equ2} and \ref{eq:Equ3} and plotted in Fig.~\ref{Fig_SingleDefects}. It is found that among the point defects the highest formation energy belongs to the Al vacancy, even in the Fe rich conditions where $\mu_{\mathrm{Fe}}=0$, it becomes 1.79 eV. That is the clear explanation why all reports and literature agreed on the observation that there will be nearly no Al vacancy and is also in agreement with F\"{a}hnle \textit{et al.} \cite{J.Mayer1999}. In Fe rich conditions the dominant defect comes to the Fe antisite with 0.14 eV, which is not surprising that making Fe antisite in Fe rich conditions is straightforward. This is  illustrated with the region starting from $\mu_{\mathrm{Fe}}=0$ up to the vertical blue line, where $\mu_{\mathrm{Fe}}\approx -0.19$ eV. In Al rich conditions the dominant defect is the Al antisite, however the defect formation energy becomes a negative value below $\mu_{\mathrm{Fe}} \approx -0.54 $ eV. Therefore, after that limit the structure is not stable anymore. In other words, there will be massive formation of antisites with the result of no B2 structure. In intermediate ranges of $\mu_{\mathrm{Fe}}$ the dominant defect is the Fe vacancy that shown between the blue and red vertical lines and the Al antisite in between the red and green vertical lines. 

\subsection{Complex Defect Formation Energies}

As listed in Fig.~\ref{Fig_ListofDefects}, the second defect group is complex defects. The formation energy of an \textit{A} type di-vacancy and di-antisite in an \textit{AB} compound can be given by:

\begin{multline}\label{eq:Equ4}
E^f_{2 vac,A} =\\
E(A_{N-2}B_N)+2 \mu_A-N E(AB),
\end{multline}

\begin{multline}\label{eq:Equ6}
E^f_{2 AS,A} = \\
E(A_{N+2}B_{N-2})+2 \mu_B-N E(AB)-2 \mu_A,
\end{multline}

\noindent where $E(A_{N-2}B_N)$ is the energy of 2 \textit{A} atoms missing supercell and $E(A_{N+2}B_{N-2})$ is the 2 \textit{A} atoms on \textit{B} sublattice structure. 

The di-vacancy and di-antisite groups can also include a defect type, where both species of atoms may contribute, namely for di-vacancy case; one \textit{A} vacancy and one \textit{B} vacancy, for the case of di-antisite; one \textit{A} antisite and one \textit{B} antisite. The corresponding defect formation energies can be formulated as: 

\begin{multline}\label{eq:Equ8}
E^f_{1 vac,A,B} = \\
E(A_{N-1}B_{N-1})+\mu_A+\mu_B-N E(AB),
\end{multline}

\begin{multline}\label{eq:Equ9}
E^f_{1 AS,A,B} = E(A_NB_N)_{\mathrm{AS}}-N E(AB),
\end{multline}

\noindent where $E(A_{N-1}B_{N-1})$ is the energy of one atom missing in both \textit{A} and \textit{B} sublattices and $E(A_NB_N)_{\mathrm{AS}}$ is the energy of one \textit{A} antisite and one \textit{B} antisite included structure.  

\begin{figure*}[t]
\begin{center}
\includegraphics[width=1.0\textwidth]{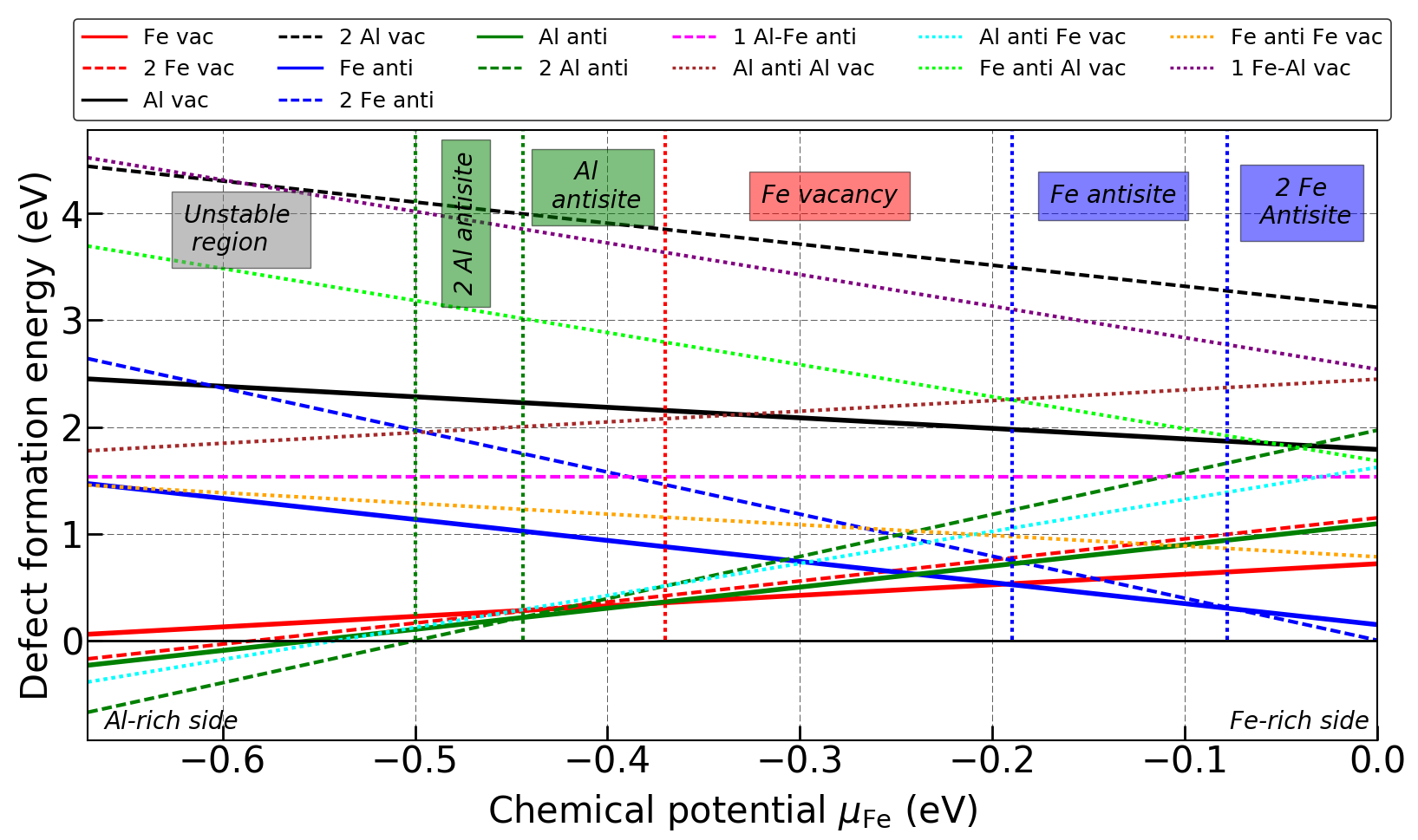}
\caption{Calculated formation energies of single and complex defects as a function of $\mu_{\mathrm{Fe}}$ for spin polarized 128 atom supercells. Single defects given by solid lines as in Fig.~\ref{Fig_SingleDefects}, di-vacancies and di-antistes given as dashed lines and 1 vacancy 1 antisite type defects given by dotted lines. The left hand side of the plot corresponds to Al rich conditions and right hand side Fe rich. The vertical lines represent the dominant defect regions.}
\label{Fig_ComplexDef}
\end{center}
\end{figure*}

The last class in complex defects involves simultaneously one vacancy and one antisite, which consist of four possible defects. Two of them only appear in the same kind of element which are: one \textit{A} vacancy one \textit{A} antisite and one \textit{B} vacancy one \textit{B} antisite. The other two involve different types of element, which are: one \textit{A} antisite one \textit{B} vacancy and one \textit{B} antisite one \textit{A} vacancy. For the same kind of elements, say one $A$ vacancy one $A$ antisite, the defect formation energy can be written as:

\begin{multline}
E^f_{1 AS,A,1 vac,A}=E(A_NB_{N-1})+\mu_B-N E(AB),
\end{multline} 

\noindent where $E(A_NB_{N-1})$ is the energy for one \textit{A} antisite and one \textit{A} vacancy. For the case of different element, say an $A$ antisite a $B$ vacancy, the defect formation energy can be formulated as:

\begin{multline}
E^f_{1 AS,A,1 vac,B}=\\
E(A_{N+1}B_{N-2})+2 \mu_B -N E(AB)-\mu_A,
\end{multline}

\noindent where $(A_{N+1}B_{N-2})$ is the energy of a structure that includes one \textit{A} antisite and one \textit{B} vacancy.

\begin{table}[h!]
\center
\small
\caption{Calculated defect formation energies (eV) with respect to pure structure reference system} 
\begin{tabular}{ l l l }
    Defect type & Fe rich cond. & Al rich cond. \\
    \hline
    Fe vacancy & 0.72 & 0.06 \\ 
    2 Fe vacancy & 1.15 & -0.17 \\
    Al vacancy & 1.79 & 2.45 \\ 
    2 Al vacancy & 3.12 & 4.44 \\
    Fe antisite & 0.15 & 1.47 \\ 
    2 Fe antisite & 0.003 & 2.64 \\
    Al antisite & 1.09 & -0.23 \\ 
    2 Al antisite & 1.97 & -0.67 \\
    Al antisite Fe antisite & 1.53 & 1.53 \\   
    Al antisite Al vacancy & 2.44 & 1.78\\ 
    Al antisite Fe vacancy & 1.62 & -0.36 \\
    Fe antisite Al vacancy & 1.68 & 3.66  \\
    Fe antisite Fe vacancy & 0.78 & 1.44  \\
     Al vacancy Fe vacancy & 2.54 & 4.52\\
    \hline
\label{Tab_ListofDefFormEne}
\end{tabular}
\end{table}

Consideration of complex defects by going beyond single defects has a significant impact on the understanding of the stability range of B2-FeAl and the dominant defect type against chemical potential of Fe. In comparison to Fig.~\ref{Fig_SingleDefects}, the dominant defect, in Fe rich conditions, is the double Fe antisite with a slightly positive value of 0.003 eV, in between $\mu_{Fe}=0$ to $\mu_{Fe} \approx-0.07$, can be seen in Fig.~\ref{Fig_ComplexDef}. Once the defect complexes are taken into account, it has been shown that the single Fe antisite is the energetically most favorable defect only between $\mu_{Fe}\approx-0.07$ eV to $\mu_{Fe} \approx-0.19$ eV. Addition of the complex defects, has not changed the dominant defect type in the intermediate regions of the chemical potential, which is the single Fe vacancy.  From $\mu_{Fe}\approx-0.37$ eV to $\mu_{Fe}\approx-0.44$ eV the dominant defect is the single Al antisite. Note that, the single Al antisite was energetically most favorable defect until $\mu_{Fe}\approx-0.54$ (check Fig.~\ref{Fig_SingleDefects}). Nevertheless, beyond   $\mu_{Fe}\approx-0.44$ the double Al antisite has a lover defect formation energy. This decreasing in the defect formation energy yields unstable B2 structure much earlier such that the lower boundary is $\mu_{\mathrm{Fe}}\approx-0.5$ eV. Therefore, the inclusion of complex defect formation energies give rise to reduction in the stability range approximately 0.04 eV. All the corresponding defect formation energies can be seen in Tab.~\ref{Tab_ListofDefFormEne} in more detail.

\section{Defect Concentrations}
\label{Sec_Def_Conc}

\begin{figure*}[t]
\begin{center}
\includegraphics[width=1.0\textwidth]{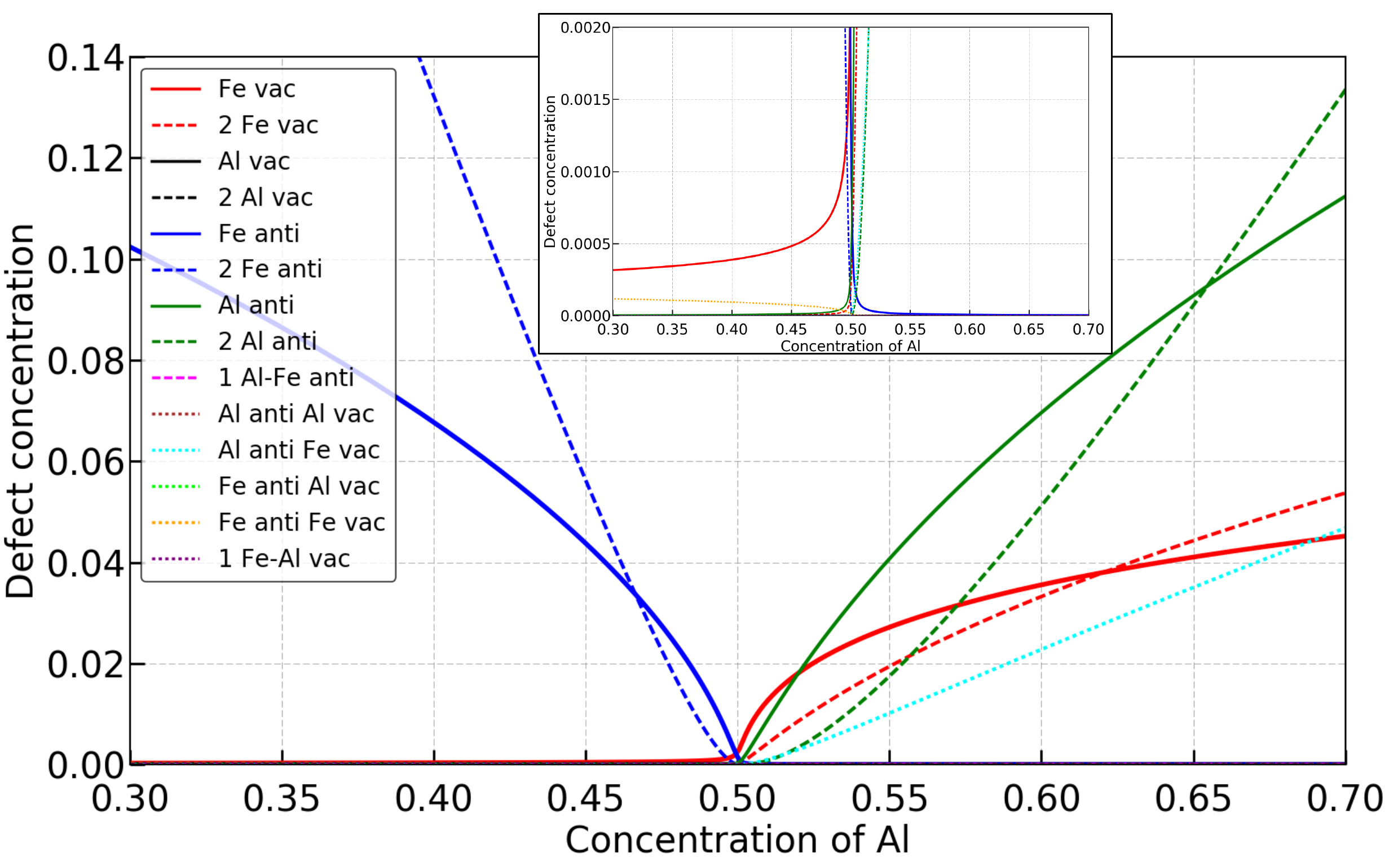}
\caption{Calculated defect concentrations of all considered defects as a function of Al concentration at 1000 K. Each concentration has been calculated by using Eq.~\ref{Eq_DefConc} and the defect formation energies that are given in Fig.~\ref{Fig_ComplexDef}. Although, results are plotted for 30-70 \% Al content, one has to notice that B2 structure is not stable slightly above the stoichiometric composition, i.e. $\sim$ 50.5 \% Al at 1000 K. The inset is a blowup of the stoichiometric composition at lower concentrations.}
\label{Fig_DefConc_1000K}
\end{center}
\end{figure*}

\begin{figure*}[t]
\begin{center}
\includegraphics[width=1.0\textwidth]{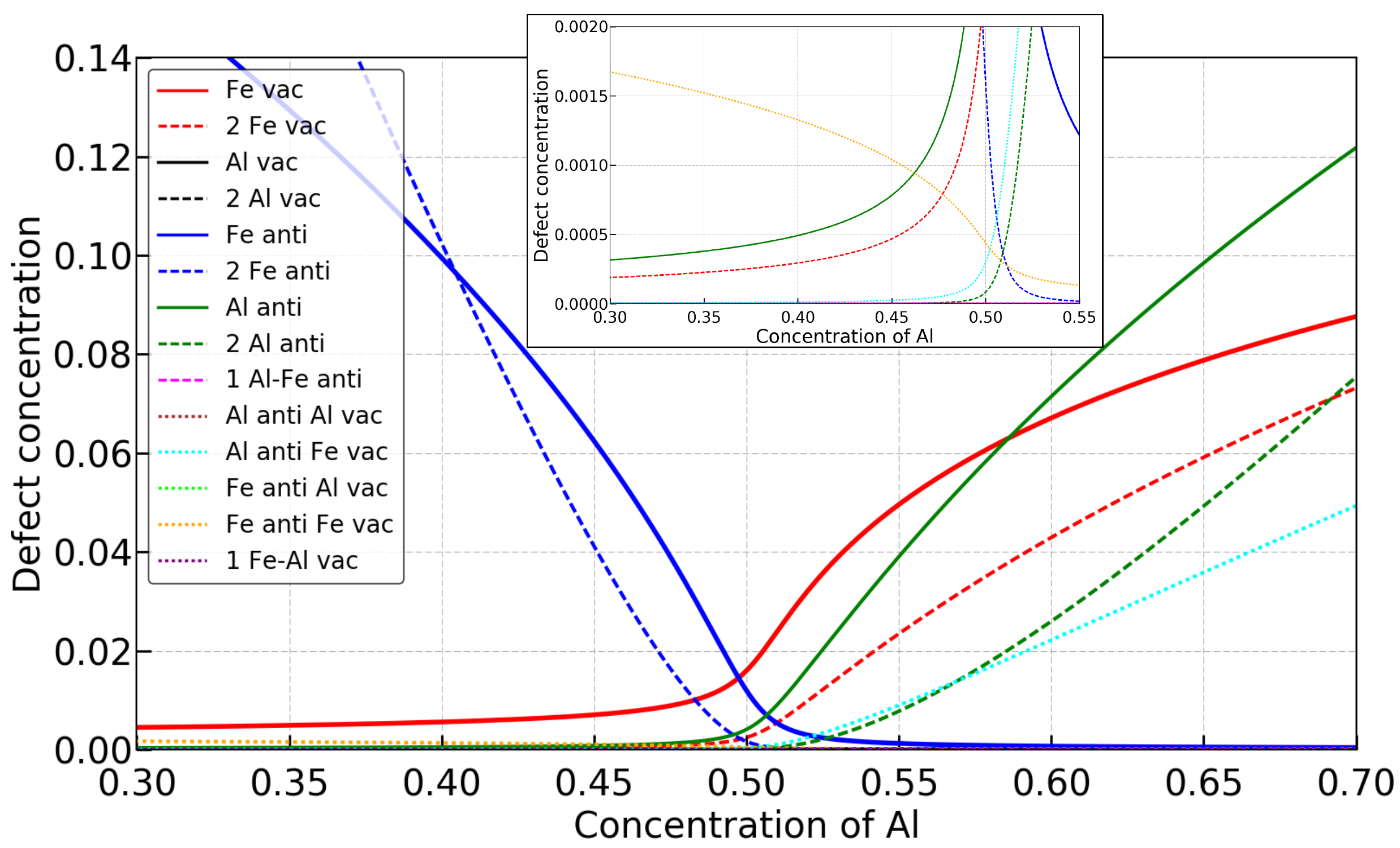}
\caption{Calculated defect concentrations of all considered defects as a function of Al concentration at 1450 K. Check the caption of Fig.~\ref{Fig_DefConc_1000K} for more details.}
\label{Fig_DefConc_1450K}
\end{center}
\end{figure*}

The defect investigations are performed in order to get quantitative expressions about the defect concentrations in B2-FeAl alloys. Once the formation energies of the defects are known, it is possible to calculate their temperature dependent concentrations. However, this is not as straightforward in an ordered binary alloy, as for unary metals, where the relation is: 

\begin{equation}
\widetilde{c} = \exp(-\frac{E_{f}}{k_{B}T}).
\label{Eq_DefConc}
\end{equation}
\noindent

Here $\widetilde{c}$ holds for the concentration, $E_{f}$ is the formation energy of the investigated defect, $k_{B}$ is the Boltzmann constant and $T$ is temperature. Mayer \textit{et al.} \cite{Mayer1995} developed a theory based on a grand canonical approach that allows to calculate the defect concentrations for B2-FeAl. The concentrations of $A$-vacancies $c_{vac,A}$ and $A$-antisites $c_{AS,A}$ are respectively given in this theory:

\begin{multline}
c_{vac,A} =\\
 \frac{e^{-(E_{vac,A}^f+\mu_A)/k_{B}T}}{1+{e^{-(E_{vac,A}^f+\mu_A)/k_{B}T}}+{e^{-(E_{AS,B}^f-\mu_B+\mu_A)/k_{B}T}}},
\label{eq:Equ13}
\end{multline}
\begin{multline}
c_{AS,A} = \\
\frac{e^{-(E_{AS,A}^f-\mu_A+\mu_B)/k_{B}T}}{1+{e^{-(E_{vac,B}^f+\mu_B)/k_{B}T}}+{e^{-(E_{AS,A}^f-\mu_A+\mu_B)/k_{B}T}}}.
\label{eq:Equ14}
\end{multline}

Nevertheless, Eq.~\ref{Eq_DefConc} gives quite similar results with Eqs.~\ref{eq:Equ13} and \ref{eq:Equ14}, especially in the Fe rich region in which we are actually most interested and also in agreement with experimental works done in 40 $\%$ and 50 $\%$ concentration of Al. It only deviates slightly above 60 $\%$ Al concentration, where the alloy is actually even not in the B2 structure but Al$_2$Fe (check the phase diagram in Ref. \cite{Sundman2009}). Therefore we stick to Eq.~\ref{Eq_DefConc} for this work.

Figs.~\ref{Fig_DefConc_1000K} and \ref{Fig_DefConc_1450K} visualize the calculated defect concentrations in B2-FeAl as a function of Al concentration. We have chosen two different temperatures: 1000 K, where B2 structure has the highest stability regime with an Al content in between 22 at.~\% to 50.5 at.~\% Al, and 1450 K, which is close to melting temperature.

Both of the figures confirm the general experimental trend \cite{Y.A.Chang1993,Pike1998,Kogachi1997,Kogachi1998,Haraguchi2002a} that the concentration of vacancies in B2-FeAl increases with increasing Al content and temperature. It is also reported in this trend that the concentration of vacancy exhibits rather gradual increases with composition in lower Al content and more rapid increase once the composition approaches to stoichiometry. The single and double Fe vacancy concentrations are indeed orders of magnitude larger than the Al vacancies. Note that, for intermediate chemical potentials, i.e. close to the B2-FeAl sotichiometry, the single Fe vacancy has the lowest formation energy of  all the defects (compare Fig.~\ref{Fig_ComplexDef}). With increasing (decreasing) Al content the formation energy decreases (increases) linearly, which immediately explains the concentration behavior in Figs.~\ref{Fig_DefConc_1000K} and \ref{Fig_DefConc_1450K}. 

There are many available experimental works in the literature that state the Fe vacancy is the dominant defect type at the stoichiometric composition \cite{Haraguchi2003,Kogachi1998,Hanc2009,Haraguchi2002a,Hanc2004}, which confirm our results. The dominant single Fe vacancy is calculated to be 0.25 \% at 1000 K and 1.6 \% at 1450 K. Earlier, it has been measured as 0.45 \% at 1073 K by dilatometry and high temperature lattice constant determination \cite{Ho1978}. Then Kogachi \cite{Kogachi1997} investigated the compositional dependence of vacancy concentration based on density and lattice constant measurements for powder samples quenched from different temperatures. He reported the vacancy concentration as $\sim$ 2.3 \% for the samples water quenched at 1073 K. It is followed by Haraguchi \cite{Haraguchi2002a} as $\sim$ 3.8 \% for the samples measured at 1250 K. One can conclude that our theoretical investigations underestimates the vacancy concentration.

However, in te case of vacancies, the biggest conflict with our results and experimental works comes on Al vacancy concentration. Experimental works (for instance check Ref.\cite{Haraguchi2002a}) showed that the vacancies occur in both sublattices, where we do not see any in Al vacancy in our results. This can be thought because of the neglecting the formation entropy in our methodology. Nevertheless, Mayer \textit{et al.} \cite{Mayer1995} found concentration of Al vacancy extremely small, of the order of 10$^{-14}$. Further he discussed to bring that concentration to considerable amounts such as 10$^{-3}$, would require a formation entropy of 25 \textit{k}$_B$ which is very unlikely. Even addition of 5 \textit{k}$_B$ in analogy would lead Al vacancy concentration of 10$^{-12}$. In agreement to our work, high defect formation energy and absence of the Al vacancies has been also reported by other theoretical works \cite{C.L.Fu1993,Mayer1995,J.Mayer1999,R.Drautz1999,J.Mayer1997,Besson1997,Ouyang2012a,Gallouze2013a}.

We predict Fe antisite as a dominant defect in the Fe rich region and increases with decreasing Al content. The reason why the antisite defects are dominant in Fe rich region can be explained by partially filled bonding states of Fe-Al, therefore with the Fe antisites in the bulk, the local Fe-Al bonding can be enhanced by \textit{d}-band filling through \textit{d-p} hybridization \cite{C.L.Fu1993}. As given in Fig.~\ref{Fig_ComplexDef}, above  $\mu_{Fe} \approx -0.19$ (corresponds to stoichiometric conditions) the single Fe antisite has the lowest defect formation energy that yields the highest defect concentration between the stoichiometirc composition and $\sim$ 47 \% Al content at 1000 K ($\sim$ 41 \% Al for 1450 K). Below that concentration the leading defect is the double Fe antiste. Experimentally, it is reported to be $\sim$ 6 \% Fe antisite at 1250 K for Fe$_{53}$Al$_{47}$ \cite{Haraguchi2002a}. In our calculations, it has been found $\sim$ 3.8 for $T$=1000 K and $\sim$ 5 \% for $T$=1450 K, which is a promising agreement. Unlike for the Al antiste the Fe antisite formation energy remains positive over the whole  $\mu_{Fe}$ (check Fig.~\ref{Fig_ComplexDef}). Therefore, we predict the B2 structure to be stable down to Al concentrations of around 30 \% for 1000 K. Since experiments \cite{Palm2007} also observe B2 to be stable above around 25 \% Al content at elevated temperatures the agreement with the theoretical results are really encouraging. 

According to available experiments and other theoretical works \cite{Haraguchi2002a,Mayer1995,C.L.Fu1993,Mayer1995,Besson1997} the Fe antisite decreases and then vanishes on the Al rich side of concentration. That completely agrees to our results. Above the stoichiometric composition, there is a competition in the Fe sublattice in between vacancy and antisite. From stoichimetric composition to $\sim$ 52 at.~\% Al the dominant defect is the single Fe vacancy then the single Al antisite takes over at 1000 K in our result. The single Fe vacancy defect dominancy range extends until 57 at.~\% 1450 K. As given in Fig.~\ref{Fig_ComplexDef}, beyond $\mu_{Fe}\approx-0.5$ the double Al antisite has a negative defect formation energy, which means the B2 structure is no longer stable. Qualitatively this observation is in agreement with experiment, where it is found that B2-FeAl becomes unstable for Al-rich conditions \cite{Palm2007} (albeit at lower Al concentrations already). Except in the region close to stoichiometry (where it is nonlinear due to the Fe vacancies) the Al antisite concentration depends on the Al content. It is also noticeable that is not negligible compared to the other defects in the Fe-rich region, especially at 1450 K. 

\section{Conclusions}
\label{Sec_Conc}

We have used \textit{ab initio} calculations based on DFT to investigate the defect structure and defect concentration in B2-FeAl. We calculated the simple bulk properties and proposed that B2 structure is ferromagnetic at 0 K. We discussed the reasons of the diversity in the previously calculated defect formation energies under three main topics which are: ($i$) supercell convergence, ($ii$) non-magnetic Fe atom considerations and ($iii$) chemical potential issue. Then we did detailed defect formation energy calculations by separating the defect types into single and complex defects. Among the single defects the Fe antiste in Fe rich region, the Fe vacancy in intermediate region, i.e in the stoichiometry, and the Al antisite in Al rich region appeared as the dominant defect types. In agreement with earlier publications we found that the Al vacancy has by far the highest formation energy over the whole concentration range. Therefore its concentration is orders of magnitude lower than for the other defects. With the inclusion of complex defects, the double defects played important role. The double Fe antisite in Fe rich conditions and the double Al antisite in Al rich condition found to be the dominant defect. The leading role of the single Fe vacancy in intermediate region remained same.

Obtained defect formation energies were used to calculate the defect concentrations at 1000 and 1450 K. We have compared well known Eq.~\ref{Eq_DefConc} and the equations developed by Mayer \textit{et al.} (Eqs.~\ref{eq:Equ13} and \ref{eq:Equ14}). Once the results are pretty similar, especially at the Fe rich region and stoichiomtric region we have used Eq.~\ref{Eq_DefConc} for rest of the work. It is found that the dominant defect in exact stoichiometry is the single Fe vacancy, which is also the dominant defect slightly above the stoichiometry for both considered temperatures.  Toward Al rich concentrations the single Al antisite takes over. Nevertheless, B2 structure is stable upto $\sim$50.5 at.\%. The leading defect type in Fe rich concentrations was found as Fe antisite and its concentration comparison to experimental data found promisingly.

Based on the achieved results, we will further extend out work on the defect kinetics and diffusion mechanism that is mediated by defects (interstitials and vacancies). This is rather important because as in the case of defect formations, there is no  general agreement on which mechanisms dominate the diffusion behavior in B2-FeAl. As a future work, we will benefit the calculated results in this work to investigate the diffusion mechanisms. This will allow us to understand the thermomechanical properties of high temperature intermetallics much better, which are closely related to both defect structure and diffusion mechanism.

\section{References}


\bibliographystyle{unsrt}
\bibliography{library.bib}

\begin{thebibliography}{10}

\bibitem{Tortorelli1992}
P.~F. Torterelli and J.~H. DeVan.
\newblock {Behavior of iron aluminides in oxidizing and oxidizing/sulfidizing
  environments}.
\newblock {\em Materials Science and Engineering}, A153:573, 1992.

\bibitem{Jordan2003}
J.~L. Jordan and S.~C. Deevi.
\newblock {Vacancy formation and effects in FeAl}.
\newblock {\em Intermetallics}, 11(6):507--528, 2003.

\bibitem{Xiao1995}
H.~Xiao and I.~Baker.
\newblock {The relationship between point defects and mechanical properties in
  Fe-Al at room temperature}.
\newblock {\em Acta metall. mater.}, 43(1):391, 1995.

\bibitem{Deevi1996}
S.C. Deevi and V.K. Sikka.
\newblock {Nickel and iron aluminides: an overview on properties, processing,
  and applications}.
\newblock {\em Intermetallics}, 4(5):357--375, 1996.

\bibitem{Liu1995}
C.T. Liu.
\newblock {Recent advances in ordered intermetallics}.
\newblock {\em Materials Chemistry and Physics}, 42(2):77--86, 1995.

\bibitem{Ho1978}
K.~Ho and R.A. Dodd.
\newblock {Point defects in FeAl}.
\newblock {\em Scripta Metallurgica}, 12(11):1055--1058, 1978.

\bibitem{Y.A.Chang1993}
D.~S.~Stone {Y. A. Chang, L. M. Pike, C. T. Liu, A. R. Bilbrey}.
\newblock {Correlation of the hardness and vacancy concentration in FeAI}.
\newblock 1:107--115, 1993.

\bibitem{Haraguchi2003}
T.~Haraguchi, K.~Yoshimi, H.~Kato, S.~Hanada, and A.~Inoue.
\newblock {Determination of density and vacancy concentration in rapidly
  solidified FeAl ribbons}.
\newblock {\em Intermetallics}, 11(7):707--711, 2003.

\bibitem{Kogachi1997}
Mineo Kogachi and Tomohide Haraguchi.
\newblock {Quenched-in vacancies in B2-structured intermetallic compound FeAl}.
\newblock {\em Materials Science and Engineering}, A230(1-2):124--131, 1997.

\bibitem{Kogachi1998}
M.~Kogachi, T.~Haraguchi, and S.~M. Kim.
\newblock {Point defect behavior in high temperature region in the B2-type
  intermetallic compound FeAl}.
\newblock {\em Intermetallics}, 6(6):499--510, 1998.

\bibitem{R.Kerl1999}
R.~Kerl, J.~Wolff, and Th. Hehenkamp.
\newblock {Equilibrium vacancy concentrations in FeAl and FeSi investigated
  with an absolute technique}.
\newblock {\em Intermetallics}, 7:301--308, 1999.

\bibitem{J.Wolff1999}
J.~Wolff, M.~Franz, A.~Broska, R.~Kerl, M.~Weinhagen, and B.~Ko.
\newblock {Point defects and their properties in FeAl and FeSi alloys}.
\newblock {\em Intermetallics}, 7:289, 1999.

\bibitem{Hanc2009}
A.~Hanc, J.~Kansy, G.~Dercz, and I.~Jendrzejewska.
\newblock {Point defect structure in B2-ordered Fe-Al alloys}.
\newblock {\em Journal of Alloys and Compounds}, 480(1):84--86, 2009.

\bibitem{Nakamura2003}
R.~Nakamura, Y.~Yamazaki, and Y.~Iijima.
\newblock {Interdiffusion in B2 Type Intermetallic Compound FeAl under High
  Pressures.}
\newblock {\em Materials Transctions}, 44(1):78, 2003.

\bibitem{Eggersmann2000}
M.~Eggersmann and H.~Mehrer.
\newblock {Diffusion in intermetallic phases of the Fe-Al system}.
\newblock {\em Philosophical Magazine A}, 80(5):1219--1244, 2000.

\bibitem{Mayer1995}
J.~Mayer, C.~Elsasser, and M.~F\"{a}hnle.
\newblock {Concentrations of Atomic Defects in B2-Fe$_x$Al$_{1-x}$}.
\newblock {\em Phys. Stat. Sol}, 191(B):283--299, 1995.

\bibitem{J.Mayer1999}
M.~F\"{a}hnle, J.~Mayer, and B.~Meyer.
\newblock {Theory of atomic defects and diffusion in ordered compounds, and
  application to B2-FeAl}.
\newblock {\em Intermetallics}, 7:315--323, 1999.

\bibitem{C.L.Fu1993}
C.~L. Fu, Y.~Y. Ye, and M.~H. Yoo.
\newblock {Equilibrium point defects in intermetallics with B2 Structure: NiAl
  and FeAl}.
\newblock {\em Phys. Rev. B}, 48(9):6712--6715, 1993.

\bibitem{R.Drautz1999}
R.~Drautz and M.~F\"{a}hnle.
\newblock {The six-jump diffusion cycle in B2 compounds}.
\newblock {\em Acta mater.}, 47(8):2437--2447, 1999.

\bibitem{Kellou2006}
A.~Kellou, T.~Grosdidier, and H.~Aourag.
\newblock {Comparative behavior of vacancy and C, B, N, O atoms single defect
  on hardening the B2-FeAl structure: An atomistic study}.
\newblock {\em Intermetallics}, 14(2):142--148, 2006.

\bibitem{Amara2010}
H.~Amara, C.~C. Fu, F.~Soisson, and P.~Maugis.
\newblock {Aluminum and vacancies in $\alpha$-iron: Dissolution, diffusion, and
  clustering}.
\newblock {\em Physical Review B}, 81(17):174101, 2010.

\bibitem{Haraguchi2005}
T.~Haraguchi, K.~Yoshimi, M.H. Yoo, H.~Kato, S.~Hanada, and A.~Inoue.
\newblock {Vacancy clustering and relaxation behavior in rapidly solidified B2
  FeAl ribbons}.
\newblock {\em Acta Materialia}, 53(13):3751--3764, 2005.

\bibitem{J.Mayer1997}
J.~Mayer and M.~F\"{a}hnle.
\newblock {On the meaning of effective formation energies, entropies and
  volumes for atomic defects in ordered compounds}.
\newblock {\em Acta mater.}, 45(5):2207--2211, 1997.

\bibitem{Besson2006}
R.~Besson, A.~Legris, and J.~Morillo.
\newblock {Influence of complex point defects in ordered alloys: An ab initio
  study of B2 Fe-Al-B}.
\newblock {\em Physical Review B}, 74(9):094103, 2006.

\bibitem{Kellou2004}
A.~Kellou, H.~I. Feraoun, T.~Grosdidier, C.~Coddet, and H.~Aourag.
\newblock {Energetics and electronic properties of vacancies, anti-sites, and
  atomic defects (B, C, and N) in B2-FeAl alloys}.
\newblock {\em Acta Materialia}, 52(11):3263--3271, 2004.

\bibitem{Murnaghan1944}
F.~D. Murnaghan.
\newblock {The compressibility of media under extreme pressures}.
\newblock {\em Proc. Nat. Acad. Sci. USA}, 30:244--247, 1944.

\bibitem{Boer1988}
F.R. de~Boer, W.C.M. Mattens, R.~Boom, A.R. Miedema, and A.K. Niessen.
\newblock {Cohesion in metals: Transition Metal Alloys}.
\newblock {\em North-Holland,Amsterdam}, 1988.

\bibitem{Villars1988}
P.~Villars and L.~D. Calvert.
\newblock {Pearson's Handbook of Crystallographic Data for intermetallic
  Phases}.
\newblock American Society for Metals, Metal Park, OH, 1988.

\bibitem{G.Kresse1996}
G.~Kresse and J.~Furthm{\"{u}}ller.
\newblock {Efficiency of ab-initio total energy calculations for metals and
  semiconductors using a plane-wave basis set}.
\newblock {\em Comput. Mat. Sci.}, 6:15--50, 1996.

\bibitem{Kresse1996}
G.~Kresse and J.~Furthm{\"{u}}ller.
\newblock {Efficient iterative schemes for ab initio total-energy calculations
  using a plane-wave basis set.}
\newblock {\em Physical review. B}, 54(16):11169--11186, 1996.

\bibitem{Perdew1996}
J.~P. Perdew, K.~Burke, and M.~Ernzerhof.
\newblock {Generalized Gradient Approximation Made Simple.}
\newblock {\em Physical review letters}, 77(18):3865--3868, 1996.

\bibitem{G.R.Caskey1972}
G.~R. Caskey, J.~M. Franz, and D.~J. Sellmyer.
\newblock {Electronic and magnetic States in Metallic Compounds-II}.
\newblock {\em J. Phys. Chem. Solids}, 34:1179--1198, 1973.

\bibitem{Mohn2001}
P.~Mohn, C.~Persson, P.~Blaha, K.~Schwarz, P.~Nov{\'{a}}k, and H.~Eschrig.
\newblock {Correlation Induced Paramagnetic Ground State in FeAl}.
\newblock {\em Physical Review Letters}, 87(19):196401, 2001.

\bibitem{Smirnov2005}
A.~Smirnov, W.~Shelton, and D.~Johnson.
\newblock {Importance of thermal disorder on the properties of alloys: Origin
  of paramagnetism and structural anomalies in bcc-based Fe$_{1-x}$Al$_x$}.
\newblock {\em Physical Review B}, 71(6):064408, 2005.

\bibitem{Zaroual2000a}
S.~Zaroual, O.~Sassi, J.~Aride, J.~Bernardini, and G.~Moya.
\newblock {Magnetic and calorimetric study of point defects in FeAl
  intermetallic compound}.
\newblock {\em Materials Science and Engineering: A}, 279(1-2):282--288, 2000.

\bibitem{Monkhorst1976}
H.~J. Monkhorst and J.~D. Pack.
\newblock {Special points for Brillouin-zone integrations}.
\newblock {\em Physical Review B}, 13(12):5188--5192, 1976.

\bibitem{Domke1984}
H.~Domke and L.K. Thomas.
\newblock {Vacancies and magnetic properties of FeAl-alloys}.
\newblock {\em Journal of Magnetism and Magnetic Materials}, 45(2-3):305--308,
  1984.

\bibitem{G.Bester2002}
G.~Bester, B.~Meyer, and M.~Fa.
\newblock {Dominant thermal defects in B2-FeAl}.
\newblock {\em Materials Science and Engineering}, A323:487--490, 2002.

\bibitem{Henkelman2006}
G.~Henkelman, A.~Arnaldsson, and H.~J{\'{o}}nsson.
\newblock {A fast and robust algorithm for Bader decomposition of charge
  density}.
\newblock {\em Computational Materials Science}, 36(3):354--360, 2006.

\bibitem{Sundman2009}
B.~Sundman, I.~Ohnuma, N.~Dupin, U.~R. Kattner, and S.~G. Fries.
\newblock {An assessment of the entire Al-Fe system including D0$_3$ ordering}.
\newblock {\em Acta Materialia}, 57(10):2896--2908, 2009.

\bibitem{Pike1998}
L.~M. Pike.
\newblock {\em {The effects of ternary alloying on the defect structure and
  mechanical properties of B2 compounds}}.
\newblock PhD thesis, University of Wisconsin-Madison, 1998.

\bibitem{Haraguchi2002a}
T.~Haraguchi and M.~Kogachi.
\newblock {Point defect behavior in B2-type intermetallic compounds}.
\newblock {\em Materials Science and Engineering: A}, 329-331:402--407, 2002.

\bibitem{Hanc2004}
A.~Hanc and J.~E. Frackowiak.
\newblock {Defect structure of Fe-Al and Fe-Al-X ( X=Ni; Cu; Cr) metallic
  powders obtained by the self-decomposition method}.
\newblock {\em Nukleonika}, 49:S7--S11, 2004.

\bibitem{Besson1997}
R.~Besson and J.~Morillo.
\newblock {Development of a semiempirical n-body noncentral potential for Fe-Al
  alloys}.
\newblock {\em Physical Review B}, 55(1):193--204, 1997.

\bibitem{Ouyang2012a}
Yifang Ouyang, Xiaofeng Tong, Chang Li, Hongmei Chen, Xiaoma Tao, Tilmann
  Hickel, and Yong Du.
\newblock {Thermodynamic and physical properties of FeAl and Fe 3Al: An
  atomistic study by EAM simulation}.
\newblock {\em Physica B: Condensed Matter}, 407(23):4530--4536, 2012.

\bibitem{Gallouze2013a}
M.~Gallouze, A.~Kellou, D.~Hamoutene, T.~Grosdidier, and M.~Drir.
\newblock {Absorption and adsorption of hydrogen in B2-FeAl: Ab initio study}.
\newblock {\em Physica B: Condensed Matter}, 416:1--7, 2013.

\bibitem{Palm2007}
F.~Stein and M.~Palm.
\newblock {Re-determination of transition temperatures in the Fe-Al system by
  differential thermal analysis}.
\newblock {\em International Journal of Materials Research}, 98:580--588, 2007.

\end{thebibliography}

\end{document}